\documentclass[journal]{IEEEtran}
\IEEEoverridecommandlockouts
\usepackage{cite}
\usepackage{amsmath,amssymb,amsfonts}
\usepackage{algorithmic}
\usepackage{graphicx}
\usepackage{textcomp}
\usepackage{xcolor}
\usepackage{multirow}
\usepackage{tabu}
\usepackage{makecell}
\usepackage{color}
\usepackage{subfigure}
\usepackage{bm}
\usepackage{balance}

\usepackage[ruled,linesnumbered]{algorithm2e}

\SetAlCapNameFnt{\scriptsize}
\SetAlCapFnt{\scriptsize}

\SetCommentSty{mycommfont}

\def\BibTeX{{\rm B\kern-.05em{\sc i\kern-.025em b}\kern-.08em
    T\kern-.1667em\lower.7ex\hbox{E}\kern-.125emX}}
    
\def\BibTeX{{\rm B\kern-.05em{\sc i\kern-.025em b}\kern-.08em T\kern-.1667em\lower.7ex\hbox{E}\kern-.125emX}}

\begin{document}

\title{LCCDE: A Decision-Based Ensemble Framework for Intrusion Detection in The Internet of Vehicles
}

\author{\IEEEauthorblockN{Li Yang$^*$, Abdallah Shami$^*$, Gary Stevens$^\dagger$, and Stephen de Rusett$^\dagger$\\}
\IEEEauthorblockA{ 
$^*$Western University, London, Ontario, Canada; e-mails: \{lyang339, abdallah.shami\}@uwo.ca \\
$^\dagger$S2E Technologies, St. Jacobs, Ontario, Canada; e-mails: \{gstevens, sderusett\}@s2etech.com \\
} 
}

\markboth{Accepted and to Appear in IEEE GlobeCom 2022}
{}

\maketitle

\begin{abstract}
Modern vehicles, including autonomous vehicles and connected vehicles, have adopted an increasing variety of functionalities through connections and communications with other vehicles, smart devices, and infrastructures. However, the growing connectivity of the Internet of Vehicles (IoV) also increases the vulnerabilities to network attacks. To protect IoV systems against cyber threats, Intrusion Detection Systems (IDSs) that can identify malicious cyber-attacks have been developed using Machine Learning (ML) approaches. To accurately detect various types of attacks in IoV networks, we propose a novel ensemble IDS framework named Leader Class and Confidence Decision Ensemble (LCCDE). It is constructed by determining the best-performing ML model among three advanced ML algorithms (XGBoost, LightGBM, and CatBoost) for every class or type of attack. The class leader models with their prediction confidence values are then utilized to make accurate decisions regarding the detection of various types of cyber-attacks. Experiments on two public IoV security datasets (Car-Hacking and CICIDS2017 datasets) demonstrate the effectiveness of the proposed LCCDE for intrusion detection on both intra-vehicle and external networks. 

\end{abstract}
\begin{IEEEkeywords}
Intrusion Detection System, Internet of Vehicles, CAN Bus, LightGBM, XGBoost, Ensemble Learning
\end{IEEEkeywords}

\section{Introduction}
With the fast development of the Internet of Things (IoT) and the Internet of Vehicles (IoV) technologies, network-controlled automobiles, such as Autonomous Vehicles (AVs) and Connected Vehicles (CVs), have begun replacing conventional vehicles \cite{iov1}. IoV systems typically consist of intra-vehicle networks (IVNs) and external networks. In IVNs, the Controller Area Network (CAN) bus is the core infrastructure enabling communication between Electronic Control Units (ECUs) to implement various functionalities \cite{iov2}. External vehicular networks, on the other hand, utilizes Vehicle-To-Everything (V2X) technology to enable the connection between smart cars and other IoV entities, such as roadside units, infrastructures, and smart devices \cite{cnnme}.

Due to the expanded network attack surfaces of IoV systems, the growing connectivity of vehicular networks has resulted in numerous security threats \cite{iov3}. In addition, there are not authentication or encryption mechanisms involved in the processing of CAN packets owing to their short length \cite{mth}. In the absence of basic security mechanisms, cybercriminals are able to insert malicious messages into IVNs and execute a variety of attacks, such as Denial of Service (DoS), fuzzy, and spoofing attacks \cite{iov3}. On the other hand, the emergence of connectivity between connected cars and external networks has made these vehicles vulnerable to a number of conventional cyber-attacks. 

\begin{figure}
     \centering
     \includegraphics[width=9.5cm]{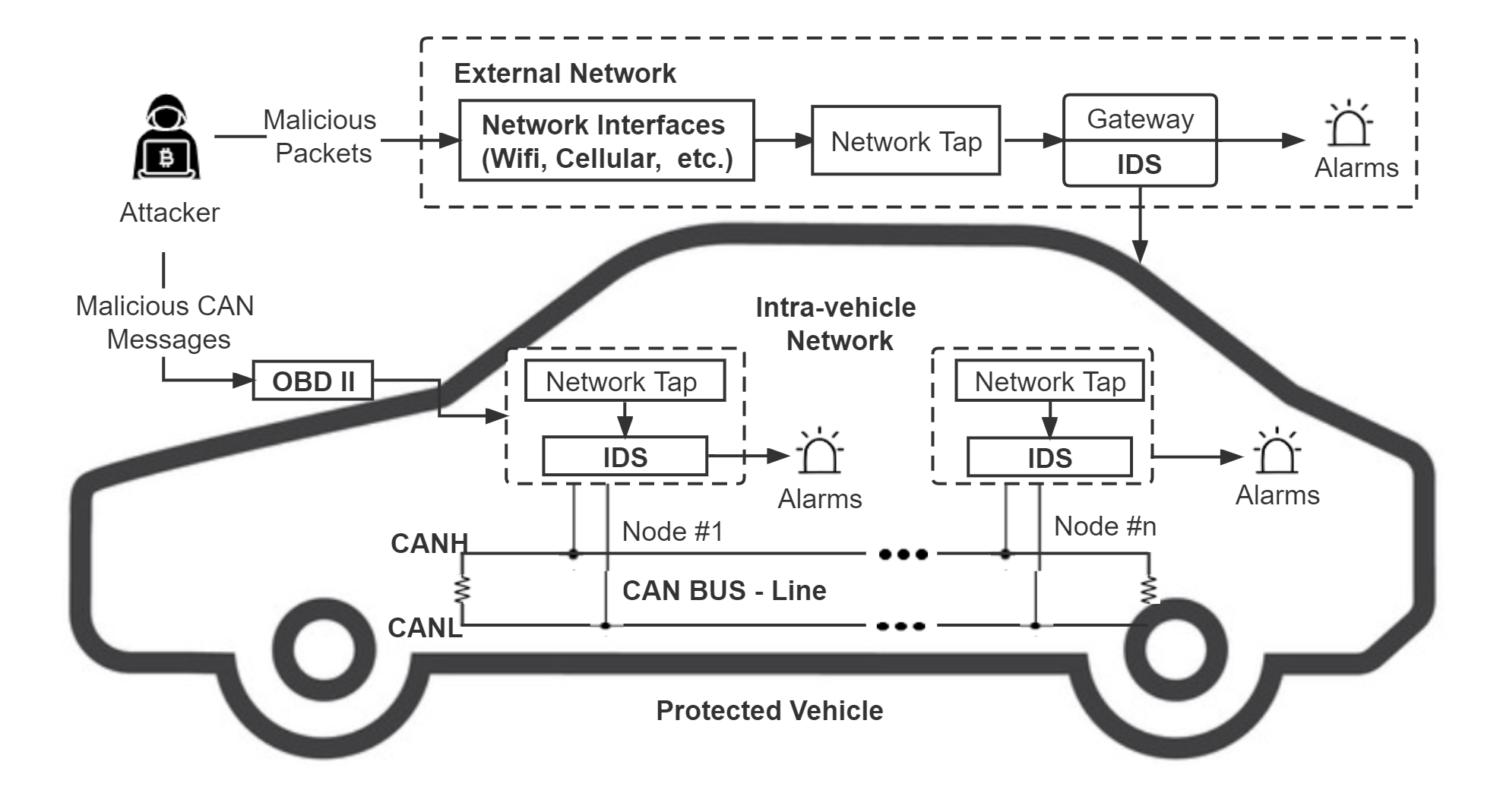}
     \caption{The IDS-protected vehicle architecture.} \label{vehicle}
\end{figure}

Figure \ref{vehicle} depicts the IoV attack scenarios, including IVN and external network attacks. Intrusion Detection Systems (IDSs) have been developed as promising solutions for detecting intrusions and defending Internet of Vehicles (IoV) systems and smart automobiles from cyber-attacks \cite{jiang}. The potential deployment of IDSs in IoV systems is also shown in Fig. \ref{vehicle}. To protect IVNs, IDSs can be placed on top of the CAN-bus to identify malicious CAN messages \cite{mth}. IDSs can also be incorporated into the gateways to detect malicious packets coming from external networks \cite{iov1}.
 
Due to the advancement of Machine Learning (ML) methods, ML-driven IDSs in IoV applications have recently drawn the interest of researchers and automotive manufacturers \cite{noorsur}. Through network traffic data analytics, ML approaches are commonly employed to construct classifier-based IDSs that can differentiate between benign network events and various cyber-attacks \cite{pwpae} \cite{tnsm}.

To apply ML models to IDS systems, it is common to observe that the prediction performance of different ML models varies significantly for different types of cyber-attack detection. Thus, a novel ensemble approach named Leader Class and Confidence Decision Ensemble (LCCDE)\footnote{
code is available at: https://github.com/Western-OC2-Lab/Intrusion-Detection-System-Using-Machine-Learning} is proposed in this paper to obtain optimal performance on all types of attack detection by integrating three advanced gradient-boosting ML algorithms, including Extreme Gradient Boosting (XGBoost) \cite{xgboost}, Light Gradient Boosting Machine (LightGBM) \cite{lightgbm}, and Categorical Boosting (CatBoost) \cite{catboost}. LCCDE aims to achieve optimal model performance by identifying the best-performing base ML model with the highest prediction confidence for each class. 

This paper mainly makes the following contributions:
\begin{enumerate}
\item It proposes a novel ensemble framework, named LCCDE, for effective intrusion detection in IoVs using class leader and confidence decision strategies, as well as gradient-boosting ML algorithms. 
\item It evaluates the proposed framework using two public IoV security datasets, Car-Hacking \cite{candata} and CICIDS2017 \cite{cicdata} datasets, representing IVN and external network data, respectively. 
\item It compares the performance of the proposed model with other state-of-the-art methods. 
\end{enumerate}

The remainder of the paper is organized as follows. Section II introduces the related work about IoV intrusion detection using ML and ensemble models. Section III presents the proposed LCCDE framework in detail. Section IV presents and discusses the experimental results. Finally, Section V concludes the paper.

\section{Related Work}
The recent surge in the number of intelligent cars has led to an increase in the development of ML models as effective solutions for IoV intrusion detection and security enhancement \cite{mj2}. Song \textit{et al.} \cite{re1} proposed a deep convolutional neural network model framework for intrusion detection in in-vehicle networks. It shows high performance on the Car-Hacking dataset. Zhao \textit{et al.} \cite{zhao} proposed an IDS framework for IoT systems based on lightweight deep neural network models. It also uses Principal Component Analysis (PCA) to reduce feature dimensionality and computational cost.

Several existing works have focused on IDS development using ensemble techniques. Yang \textit{et al.} \cite{treeme} proposed a stacking ensemble framework for network intrusion detection in IoV systems using tree-based ML models. The stacking ensemble model shows high accuracy on the CAN-Intrusion and CICIDS2017 datasets. Elmasry \textit{et al.} \cite{re2} proposed an ensemble model for network intrusion detection using three deep learning models: Deep Neural Networks (DNN), Long Short-Term Memory (LSTM), and Deep Belief Networks (DBN). Chen \textit{et al.} \cite{chen} proposed a novel ensemble IDS framework, named All Predict Wisest Decides (APWD), to detect intrusions and make decisions based on the wisest model for each class. However, it only achieves an accuracy of 79.7\% on the NSL-KDD dataset. 

Although many of the related works achieve high performance in intrusion detection tasks of IoV systems, there is still much room for performance improvement. Existing IDS frameworks can be improved with the use of more advanced ML algorithms and ensemble strategies. To the best of our knowledge, our proposed LCCDE framework is the first technique that leverages both leader class and prediction confidence strategies to construct ensemble IDSs. The use of three advanced gradient-boosting algorithms also improves the effectiveness of intrusion detection.

\section{Proposed Framework}
\subsection{System Overview}

\begin{figure*}
     \centering
     \includegraphics[width=15.5cm]{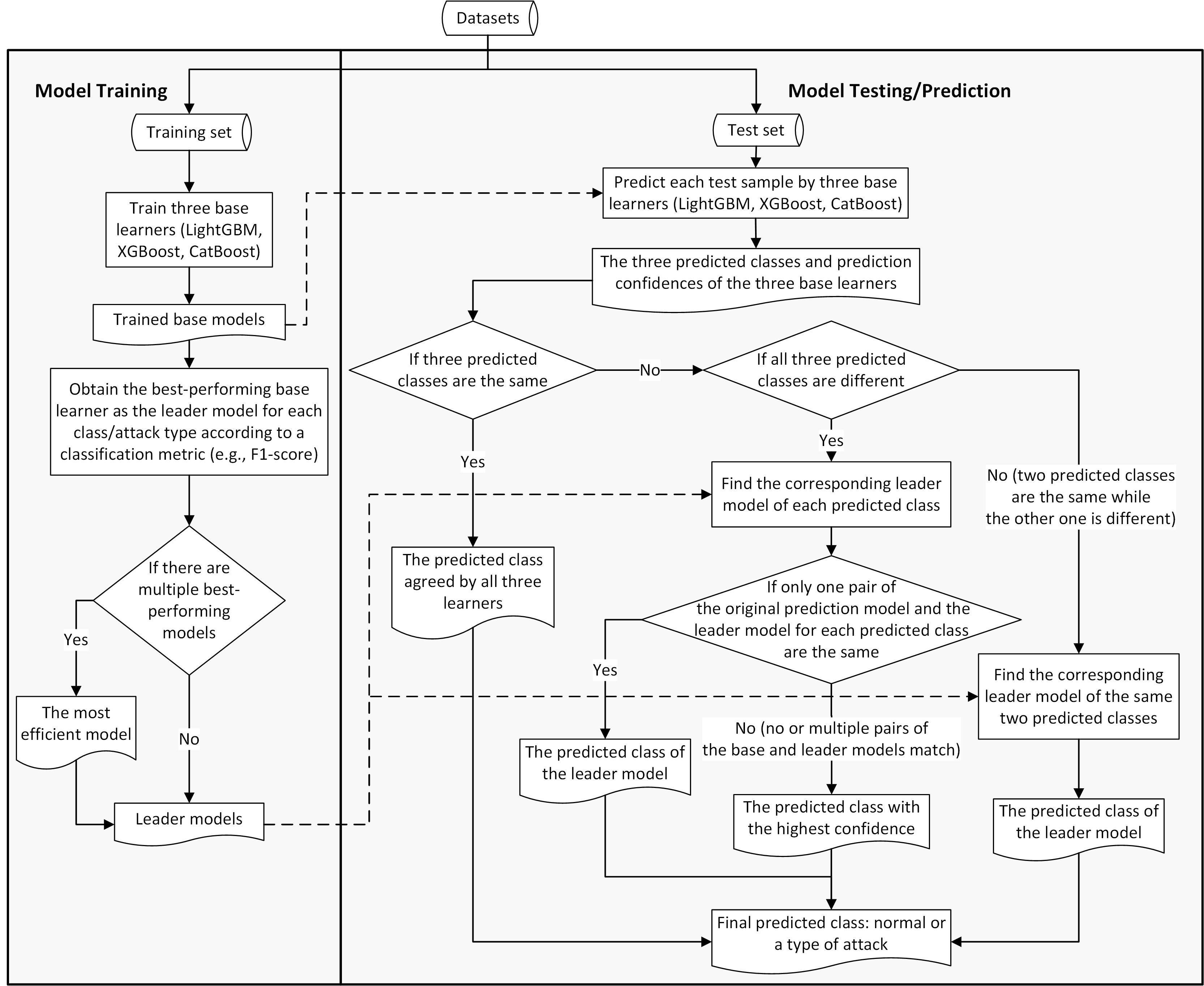}
     \caption{The framework of the proposed LCCDE model. } \label{frame}
\end{figure*}

The purpose of this work is to develop an ensemble IDS framework that can effectively detect various types of attacks on both IVN and external vehicular networks. Figure \ref{frame} demonstrates the overall framework of the proposed system, consisting of two phases: model training and model prediction. At the model training stage, three advanced ML algorithms, XGBoost \cite{xgboost}, LightGBM \cite{lightgbm}, and CatBoost \cite{catboost}, are trained on the IoV traffic dataset to obtain the leader models for all classes/types of attacks. At the model prediction stage, the class leader models and their prediction confidences are used to accurately detect attacks. The algorithm details are provided in this section. 

\subsection{Base Machine Learning Models}
A decision tree (DT) is a basic ML algorithm that uses a tree structure to make decisions based on the divide and conquer technique \cite{treeme}. In DTs, the decision nodes represent the decision tests, while the leaves indicate the result classes. Gradient Boosting Decision Tree (GBDT) is an iterative DT algorithm that constructs multiple DTs and aggregates their prediction outputs. To improve the performance of basic GBDTs, three advanced gradient-boosting algorithms, XGBoost \cite{xgboost}, LightGBM \cite{lightgbm}, and CatBoost \cite{catboost}, have been developed and widely used in many applications. These three gradient-boosting algorithms are used in the proposed system to build the LCCDE ensemble framework. 

XGBoost is a popular gradient-boosting DT algorithm designed for the speed and performance improvement of GBDTs \cite{xgboost}. XGBoost uses a regularization term and a Second-Order Taylor Approximation for the summation of the squared errors to minimize the loss function and reduce over-fitting. XGBoost has a low computational complexity of $O(K d\|\mathbf{x}\| \log n)$, where $K$ is the number of trees, $d$ is the maximum tree depth, $\|\mathbf{x}\|$ is the number and non-missing samples, and $n$ is the data size \cite{xgboost}. Additionally, XGBoost support parallel execution to save model learning time. 

LightGBM is a fast and robust ensemble ML model constructed by multiple DTs \cite{lightgbm}. LightGBM's key advantage over other ML methods is its capacity to efficiently handle large-scale and high-dimensional data. Gradient-based One-Side Sampling (GOSS) and Exclusive Feature Bundling (EFB) are the two core strategies of LightGBM \cite{lightgbm}. GOSS is a down-sampling method that only preserves data samples with large gradients and randomly discards small gradient samples to accelerate model training and reduce memory consumption. EFB is a feature engineering method that regroups mutually exclusive features into bundles as single features to minimize feature size and improve model training efficiency. By employing GOSS and EFB, the data size can be reduced significantly without the loss of critical information. The time and space complexity of LightGBM has also been reduced to $O(N^\prime F^\prime)$, where $N^\prime$ is the reduced number of samples after using GOSS, $F^\prime$ is the bundled number of features after employing EFB \cite{iotm}.

CatBoost is another advanced gradient-boosting algorithm designed to process categorical features more effectively \cite{catboost}. CatBoost, in comparison to existing gradient-boosting models, includes three significant model enhancement components: symmetric trees, ordered boosting, and native feature support. In symmetric trees, leaves are split under the same condition as in prior trees, and the pair of feature splits with the lowest loss is applied to all nodes. Using symmetric trees can improve model prediction speed and reduce over-fitting. Ordered boosting is a permutation-driven technique that prevents overfitting on small datasets by training a model on a subset while calculating residuals on another subset. CatBoost's native feature support indicates that it can directly process all types of features, such as numerical, textual, and categorical features, without the need for extra pre-processing. CatBoost is an ensemble model with low computational complexity of $O(SN)$, where $S$ is the number of permutations of the subsets and $N$ is the number of base DT models \cite{catboost}.

The primary reasons for selecting XGBoost, LightGBM, and CatBoost as base learners are as follows \cite{xgboost} - \cite{catboost}: 
\begin{enumerate}
\item These three ML models are all robust ensemble models that have had great success in a variety of data analytics applications \cite{hpo}. 
\item These three ML models can automatically generate feature importance scores and select features during their training process, which saves time and resources by avoiding the need for extra feature engineering. 
\item These three ML models are fast models with relatively low computational complexity. Additionally, they all support parallelization and Graphics Processing Unit (GPU) execution, which can further improve model learning speed. 
\item These three ML models include randomness in their model construction process, enabling people to develop a robust ensemble model with high diversity and generalizability. 
\end{enumerate}

\subsection{LCCDE: Proposed Ensemble Algorithm}

\begin{algorithm}[t]
    {\scriptsize
	\caption{Leader Class and Confidence Decision Ensemble (LCCDE) - Model Training}
	\label{train}
	\LinesNumbered
	\KwIn{
	\\\quad $D_{train}$: the training set,
	\\\quad $M = \{M_1,M_2,M_3\}$: the base ML model list, including $M_1$ = LightGBM, $M_2$ = XGBoost, $M_3$ = CatBoost,
	\\\quad $c=1,2,\dots,n$: the class list for $n$ different classes.
	}
	\KwOut{
	\\\quad $M = \{M_1,M_2,M_3\}$: the trained base model list, 
	\\\quad $LM = \{LM_1,LM_2,\dots,LM_n\}$: the leader model list for all classes.}

    $M_1 \leftarrow Training(M_1,D_{train})$; \tcp*[f]{Train the LightGBM model}\\
    $M_2 \leftarrow Training(M_2,D_{train})$; \tcp*[f]{Train the XGBoost model}\\
    $M_3 \leftarrow Training(M_3,D_{train})$; \tcp*[f]{Train the CatBoost model}\\

	\For(\tcp*[f]{For each class (normal or a type of attack), find the leader model}){$c=1,2,\dots,n$}{	    $Mlist_c \leftarrow BestPerforming(M_1,M_2,M_3,c)$; \tcp*[f]{Find the best-performing model for each class (\textit{e.g.,} has the highest F1-score)}\\
	
    	\uIf(\tcp*[f]{If only one model has the highest F1 }){$Len(Mlist_c)==1$ } 
        	{
        	$LM_c\leftarrow Mlist_c[0]$; \tcp*[f]{Save this model as the leader model for the class $c$}\\
        	}

    	\Else (\tcp*[f]{If multiple ML models have the same highest F1-score}) {
    	    $LM_c\leftarrow MostEfficient(Mlist_c)$; \tcp*[f]{Save the fastest or most efficient model as the leader model for the class $c$}\\
    	
    	    }
        $LM\leftarrow LM\cup\{LM_c\}$;\tcp*[f]{Collect the leader model for each class}
    }
}
\end{algorithm}

\begin{algorithm}[t]
    {\scriptsize
	\caption{Leader Class and Confidence Decision Ensemble (LCCDE) - Model Prediction}
	\label{test}
	\LinesNumbered
	\KwIn{
	\\\quad $D_{test}$: the test set,
	\\\quad $M = \{M_1,M_2,M_3\}$: the trained base ML model list, including $M_1$ = LightGBM, $M_2$ = XGBoost, $M_3$ = CatBoost,
	\\\quad $c=1,2,\dots,n$: the class list for $n$ different classes.
	}
	\KwOut{
	\\\quad $L_{test}$: the prediction classes for all test samples in $D_{test}$.}
	
	\For(\tcp*[f]{For each test sample}){\rm{each data sample} $x_i\in D_{test}$}{	
	$L_{i1}, p_{i1} \leftarrow Prediction\left(M_{1}, x_{i}\right)$; \tcp*[f]{Use the trained LightGBM model to predict the sample, and save the predicted class \& confidence}\\
	$L_{i2}, p_{i2} \leftarrow Prediction\left(M_{2}, x_{i}\right)$; \tcp*[f]{Use XGBoost to predict}\\
	$L_{i3}, p_{i3} \leftarrow Prediction\left(M_{3}, x_{i}\right)$; \tcp*[f]{Use CatBoost to predict}\\

    	\uIf(\tcp*[f]{If the predicted classes of all the three models are the same }){$L_{i1}==L_{i2}==L_{i3}$} 
        	{
        	$L_{i} \leftarrow L_{i1}$; \tcp*[f]{Use this predicted class as the final predicted class}\\
        	}
 	    
 	    \uElseIf(\tcp*[f]{If the predicted classes of all the three models are different}){$L_{i1}!=L_{i2}!=L_{i3}$}
	        {
            \For(\tcp*[f]{For each prediction model}){$j=1,2,3$}{
                \If(\tcp*[f]{Check if the predicted class’s original ML model is the same as its leader model}){$M_{j}==LM_{L_{i,j}}$} 
                    {
                    $L\_list_{i}\leftarrow L\_list_{i}\cup\{L_{i,j}\}$; \tcp*[f]{Save the predicted class}\\
                    $p\_list_{i}\leftarrow p\_list_{i}\cup\{p_{i,j}\}$; \tcp*[f]{Save the confidence}\\                    
                    }
                }
        	\uIf(\tcp*[f]{If only one pair of the original model and the leader model for each predicted class is the same}){$Len(L\_list_{i})==1$} 
            	{
            	$L_j \leftarrow L\_list_i[0]$; \tcp*[f]{Use the predicted class of the leader model as the final prediction class}\\
            	}
        	\Else (\tcp*[f]{If no pair or multiple pairs of the original prediction model and the leader model for each predicted class are the same}) {
            	\If{$Len(L\_list_{i})==0$} 
            	    {
            	    $p\_list_{i} \leftarrow \{p_{i1},p_{i2},p_{i3}\}$; \tcp*[f]{Avoid empty probability list}\\
            	    }
            	$p\_max_{i} \leftarrow \max(p\_list_{i})$; \tcp*[f]{Find the highest confidence}\\
	        	\uIf(\tcp*[f]{Use the predicted class with the highest confidence as the final prediction class}){$p\_max_{i}==p_{i1}$} 
	        	    {
	            	$L_{i} \leftarrow L_{i1}$; \\
	        	    }
	     	    \uElseIf{$p\_max_{i}==p_{i2}$}
	     	        {
	     	        $L_{i} \leftarrow L_{i2}$; \\
	     	        }
        	    \Else 
        	        {
        	        $L_{i} \leftarrow L_{i3}$; \\
        	        }
        	    
        	    }

	        }

    	\Else (\tcp*[f]{If two predicted classes are the same and the other one is different}) {
            $n \leftarrow mode(L_{i1},L_{i2},L_{i3})$; \tcp*[f]{Find the predicted class with the majority vote}\\
            $L_i \leftarrow Prediction(M_{n},x_{i})$; \tcp*[f]{Use the predicted class of the leader model as the final prediction class}\\    	    
    	    }
        $L_{test}\leftarrow L_{test}\cup\{L_i\}$; \tcp*[f]{Save the predicted classes for all tested samples};
    }
}
\end{algorithm}

The performance of different ML models often varies on different types of attack detection tasks. For example, when applying multiple ML models on the same network traffic dataset, a ML model perform the best for detecting the first type of attack (\textit{e.g.,} DoS attacks), while another ML model may outperform other models for detecting the second type of attack (\textit{e.g.,} sniffing attacks). Therefore, this work aims to propose an ensemble framework that can achieve optimal model performance for the detection of every type of attack. Ensemble learning is a technique that combines multiple base ML models to improve learning performance and generalizability \cite{cnnme}. The proposed ensemble model is constructed using XGBoost, LightGBM, and CatBoost, three advanced gradient-boosting ML methods introduced in Section III-B. 

Figure \ref{frame} demonstrates the process of the proposed LCCDE framework in two phases: model training and model prediction. The detailed procedures of the training and prediction phases are also described in Algorithms \ref{train} \& \ref{test}, respectively. At the training stage, the LCCDE framework aims to obtain leader models for all classes via the following steps:
\begin{enumerate}
\item \textit{Train three base learners}. The three base ML models (XGBoost, LightGBM, and CatBoost) are trained on the training set to obtain base learners.
\item \textit{Evaluate base learners}. The performance of the three ML models for each class (normal or a type of attack) is evaluated using cross-validation and F1-scores. F1-scores are chosen because it is a comprehensive performance metric and works well with imbalanced datasets.
\item \textit{Determine the leader model for each class}. For each class, the best-performing ML model with the highest F1-score is selected as the leader model for each class. If multiple top-performing ML models have identical highest F1-scores, the most efficient ML model with the highest speed is chosen as the final leader model. 
\end{enumerate}

\begin{table*}[!t]
\centering%
\caption{Model Performance Comparison for Each Class in The Two Datasets}
\setlength\extrarowheight{1pt}
\scalebox{0.80}{
\begin{tabular}{|>{\centering\arraybackslash}m{6.5em}|>{\centering\arraybackslash}m{4.5em}|>{\centering\arraybackslash}m{4.3em}|>{\centering\arraybackslash}m{4.3em}|>{\centering\arraybackslash}m{4.3em}|>{\centering\arraybackslash}m{4.3em}|>{\centering\arraybackslash}m{4.5em}|>{\centering\arraybackslash}m{4.3em}|>{\centering\arraybackslash}m{4.5em}|>{\centering\arraybackslash}m{4.3em}|>{\centering\arraybackslash}m{4.3em}|>{\centering\arraybackslash}m{4.5em}|>{\centering\arraybackslash}m{4.5em}|}
\hline
\multirow{5}{*}{\textbf{ Method }} & \multicolumn{5}{c!{\color{black}\vrule}}{\textbf{ Car-Hacking Dataset }}                                                                                                                                        & \multicolumn{7}{c!{\color{black}\vrule}}{\textbf{ CICIDS2017 Dataset}}                                                                                                                                                                                                                                                       \\ 
\cline{2-13}
                                   & \textbf{ F1 (\%) of Class 1: Normal  } & \textbf{ F1 (\%) of Class 2: DoS } & \textbf{ F1 (\%) of Class 3: Fuzzy } & \textbf{ F1 (\%) of Class 4: Gear Spoofing } & \textbf{ F1 (\%) of Class 5: RPM Spoofing } & \textbf{ F1 (\%) of Class 1: Normal  } & \textbf{ F1 (\%) of Class 2: DoS } & \textbf{ F1 (\%) of Class 3: Sniffing } & \textbf{ F1 (\%) of Class 4: Brute-Force } & \textbf{ F1 (\%) of Class 5: Web Attack } & \textbf{ F1 (\%) of Class 6: Botnets } & \textbf{F1 (\%) of Class 7: Infiltration}  \\ 
\hline
LightGBM \cite{lightgbm}                      & \textbf{99.9998 }                     & \textbf{100.0 }                   & \textbf{99.995 }                    & \textbf{100.0 }                             & \textbf{100.0 }                            & 99.863                      & \textbf{100.0 }                   & \textbf{99.889 }                       & 99.222                                     & \textbf{99.354 }                         & \textbf{100.0 }            & \textbf{85.714 }                 \\ 
\hline
XGBoost \cite{xgboost}                        & 99.9996                                & \textbf{100.0 }                   & 99.990                               & \textbf{100.0 }                             & \textbf{100.0 }                            & 99.863                      & \textbf{100.0 }                   & \textbf{99.889 }                       & \textbf{99.351 }                          & 99.137                                    & \textbf{100.0 }            & \textbf{85.714 }                 \\ 
\hline
CatBoost \cite{catboost}                      & 99.9996                                & \textbf{100.0 }                   & 99.990                               & \textbf{100.0 }                             & \textbf{100.0 }                            & 99.794                                 & 99.754                             & 99.557                                  & 99.094                                     & \textbf{99.354 }                         & \textbf{100.0 }            & \textbf{85.714 }                 \\ 
\hline
\textbf{ Proposed LCCDE }          & \textbf{99.9998 }                     & \textbf{100.0 }                   & \textbf{99.995 }                    & \textbf{100.0 }                             & \textbf{100.0 }                            & \textbf{99.876 }                      & \textbf{100.0 }                   & \textbf{99.889 }                       & \textbf{99.351 }                          & \textbf{99.354 }                         & \textbf{100.0 }            & \textbf{85.714 }                 \\
\hline
\end{tabular}
}
\label{f1}
\end{table*}

After the training process, the trained leader models for all classes are utilized for model prediction. At the model prediction stage, the LCCDE framework predicts each test sample based on the following steps:
\begin{enumerate}
\item \textit{Make initial predictions}. The three trained base ML models obtained from the training process are used to make initial predictions. Their predicted classes and the corresponding prediction confidences are retained for further analysis. Confidence is a probability value used to quantify how confident the model is about its predictions. 
\item \textit{Check if the three predicted classes are the same.} If they are the same, the predicted class agreed by all three base learners is used as the final predicted class.
\item \textit{Check if the three predicted classes are different.} If they are all different, the corresponding leader model for each predicted class is compared to the base learner that has predicted this class. If only one pair of the leader and base models match, their predicted class is used as the final predicted class; otherwise, the predicted class with the highest prediction confidence is used as the final predicted class.
\item \textit{Check if two predicted classes are the same and the other one is different.} If so, the corresponding leader model of the same two predicted classes is used to make the final prediction. 
\end{enumerate}

In brief, LCCDE detects attacks based on the following three principles:
\begin{enumerate}
\item It uses the trained ML models to generate initially predicted classes.
\item It uses the leader models for each class to make the final predictions.
\item If there are multiple leader models for different classes, it selects the leader model with the highest prediction confidence to make final decisions. 
\end{enumerate}	
	
The computational complexity of LCCDE is $O(NCK)$, where $N$ is the data size, $C$ is the number of classes, $K$ is the complexity of base models. Thus, its complexity mainly depends on the complexity of all base ML models. In the proposed framework, three fast gradient-boosting ML algorithms are used to achieve low overall complexity. These three algorithms can be replaced by other ML algorithms using the same generic LCCDE strategy according to specific tasks. LCCDE is designed to address the difficult samples that cannot be correctly predicted by individual ML models. By using LCCDE, the final ensemble model can achieve optimal performance for detecting every type of attack.

\section{Performance Evaluation}
\subsection{Experimental Setup}
To develop the proposed IDS, the models were implemented using Scikit-learn, Xgboost \cite{xgboost}, Lightgbm\cite{lightgbm}, and Catboost\cite{catboost} libraries in Python. The experiments were conducted on a Dell Precision 3630 computer with an i7-8700 processor and 16 GB of memory, representing an IoV server machine.

The proposed LCCDE framework is evaluated on two public benchmark IoV network security datasets, Car-Hacking \cite{candata} and CICIDS2017 \cite{cicdata} datasets, representing the IVN and external network data, respectively. The Car-Hacking dataset \cite{candata} is created by transmitting CAN messages into a real vehicle’s CAN bus. It has nine features (\textit{i.e., }CAN ID and the eight bits of the CAN message data field) and four types of attacks (\textit{i.e., }DoS, fuzzy, gear spoofing, and Revolutions Per Minute (RPM) spoofing attacks). The CICIDS2017 dataset \cite{cicdata} is a state-of-the-art general cyber-security dataset including the most updated types of attacks (\textit{i.e., }DoS, sniffing, brute-force, web-attacks, botnets, and infiltration attacks).

To evaluate the proposed LCCDE model, five-fold cross-validation is used in the training process to select leader class models, and 80\%/20\% hold-out validation is then used in the testing process to evaluate the model on the unseen test set. As network traffic data is often highly imbalanced and contains only a small proportion of attack samples, four performance measures, including accuracy, precision, recall, and F1-scores, are utilized to evaluate the model performance \cite{cnnme}. The execution time, including the model training and test time, is used to evaluate the efficiency of the model.

\subsection{Experimental Results and Discussion}

\begin{table}[!t]
\caption{Performance Evaluation of Models on Car-hacking Dataset}
\centering
\setlength\extrarowheight{1pt}
\scalebox{0.82}{
\begin{tabular}{|>{\centering\arraybackslash}p{7em}|>{\centering\arraybackslash}p{4.5em}|>{\centering\arraybackslash}p{4.0em}|>{\centering\arraybackslash}p{3.5em}|>{\centering\arraybackslash}p{3.5em}|>{\centering\arraybackslash}p{4.5em}|}
\hline
\textbf{Method}         & \textbf{Accuracy (\%)} & \textbf{Precision (\%)} & \textbf{Recall (\%)} & \textbf{F1 (\%)} & \textbf{Execution Time (s)}  \\ 
\hline
KNN \cite{cancom}                  & 97.4                     & 96.3                      & 98.2                   & 93.4               & 195.6                          \\ 
\hline
SVM \cite{cancom}                  & 96.5                     & 95.7                      & 98.3                   & 93.3               & 1345.3                         \\ 
\hline
LSTM-AE \cite{lstm}              & 99.0                     & 99.0                      & 99.9                   & 99.0               & -                              \\ 
\hline
DCNN \cite{re1}                 & 99.93                    & 99.84                     & 99.84                  & 99.91              & -                              \\ 
\hline
LightGBM \cite{lightgbm}             & 99.9997                  & 99.9997                   & 99.9997                & 99.9997            & 10.7                           \\ 
\hline
XGBoost \cite{xgboost}               & 99.9994                  & 99.9994                   & 99.9994                & 99.9994            & 45.3                           \\ 
\hline
CatBoost \cite{catboost}             & 99.9994                  & 99.9994                   & 99.9994                & 99.9994            & 88.6                           \\ 
\hline
\textbf{Proposed LCCDE} & \textbf{99.9997}                  & \textbf{99.9997}                   & \textbf{99.9997}                & \textbf{99.9997}            & 185.1                          \\
\hline
\end{tabular}
}
\label{can}%
\end{table}

\begin{table}[!t]
\caption{Performance Evaluation of Models on CICIDS2017}
\centering
\setlength\extrarowheight{1pt}
\scalebox{0.82}{
\begin{tabular}{|>{\centering\arraybackslash}p{7em}|>{\centering\arraybackslash}p{4.5em}|>{\centering\arraybackslash}p{4.0em}|>{\centering\arraybackslash}p{3.5em}|>{\centering\arraybackslash}p{3.5em}|>{\centering\arraybackslash}p{4.5em}|}
\hline
\textbf{Method}         & \textbf{Accuracy (\%)} & \textbf{Precision (\%)} & \textbf{Recall (\%)} & \textbf{F1 (\%)} & \textbf{Execution Time (s)}  \\ 
\hline
KNN \cite{cicdata}                  & 96.3              & 96.2              & 93.7              & 96.3              & 1558.3  \\ 
\hline
RF \cite{cicdata}                   & 98.82             & 98.8              & 99.955            & 98.8              & 135.1   \\ 
\hline
DBN \cite{re2}                  & 98.95             & 95.82             & 95.81             & 95.81             & -       \\ 
\hline
Stacking \cite{treeme}             & 99.80             & 99.75             & 99.89             & 99.70             & 278.6   \\ 
\hline
LightGBM \cite{lightgbm}             & 99.794            & 99.795            & 99.794            & 99.792            & 14.3    \\ 
\hline
XGBoost \cite{xgboost}               & 99.794            & 99.795            & 99.794            & 99.792            & 44.7    \\ 
\hline
CatBoost \cite{catboost}             & 99.683            & 99.684            & 99.683            & 99.680            & 73.7    \\ 
\hline
\textbf{Proposed LCCDE} & \textbf{99.813} & \textbf{99.814} & \textbf{99.913} & \textbf{99.811} & 168.9   \\
\hline
\end{tabular}
}
\label{cic}%
\end{table}

The experimental results of evaluating the three base ML models (LightGBM, XGBoost, CatBoost) and the proposed LCCDE model on the Car-Hacking and CICIDS2017 datasets are shown in Tables \ref{f1} – \ref{cic}. Table \ref{f1} illustrates the performance of the four models for detecting every type of attack in the two datasets based on their F1-scores. It is noticeable that the F1-scores of different base ML models vary for different types of attack detection. For example, on the CICIDS2017 dataset, LightGBM achieves the highest F1-score among the three base learners for detecting normal samples, DoS, sniffing, webattacks, botnets, and infiltration attacks, while XGBoost outperforms LightGBM for the brute-force attack detection. As shown in Table \ref{f1}, the proposed LCCDE ensemble model can achieve the highest F1-score for every class. Thus, as shown in Tables \ref{can} and \ref{cic}, the overall F1-scores of the proposed model are also the highest among the four utilized ML models on the two datasets. The proposed LCCDE model achieves a near-perfect F1-score on the Car-Hacking dataset (99.9997\%), and improved its F1-score from 99.792\% to 99.811\% on the CICIDS2017 dataset. This demonstrates the benefits of identifying the best-performing base models for each class to construct the LCCDE ensemble model.

Tables \ref{can} and \ref{cic} also compare the performance of the proposed technique with existing state-of-the-art methods \cite{cicdata} \cite{re1} \cite{treeme} \cite{re2} \cite{cancom} \cite{lstm} on the two datasets. The proposed LCCDE model outperforms other methods by at least 0.09\% and 0.11\% F1-score improvements on the Car-Hacking and CICIDS2017 datasets, respectively. As an ensemble approach, the proposed LCCDE model has a longer execution time than the other three base gradient-boosting models, but it is still faster than many other ML algorithms, such as K-Nearest Neighbors (KNN) and Support Vector Machine (SVM). This is because the proposed ensemble model is built using low complexity ML models with parallel execution and GPU support. To summarize, the proposed model can achieve the highest F1-scores among the compared methods with relatively low execution time on the two benchmark datasets.

\section{Conclusion}
For the purpose of enhancing IoV security, Machine Learning (ML) algorithms have been used as promising solutions to detect various types of cyber-attacks. However, ML models often perform differently for different types of attack detection. To achieve optimal performance on all types of attack detection in IoV networks, a novel ensemble method, namely Leader Class and Confidence Decision Ensemble (LCCDE), is proposed in this paper. It identifies the best-performing ML models for each type of attack detection as the leader class models to construct a robust ensemble model. Additionally, the prediction confidence information is utilized to help determine the final prediction classes. Three advanced gradient-boosting ML algorithms, XGBoost, LightGBM, and CatBoost, are utilized to construct the proposed LCCDE ensemble model due to their high effectiveness and efficiency. Through the experiments, the proposed IDS framework achieves high F1-scores of 99.9997\% and 99.811\% on the Car-Hacking and CICIDS2017 datasets, representing intra-vehicle and external vehicular network data, respectively. Moreover, the proposed model’s F1-scores are higher than other compared ML methods for detecting every type of attack. This illustrates the benefits of the proposed leader class-based strategy.

\section{Acknowledgement}
This work is partially supported by The Canadian Urban Transit Research \& Innovation Consortium (CUTRIC).

\end{document}